\documentstyle[12pt]{article}
\begin{document}

% Accents and foreign (in text):

\let\und=\b         % bar-under (but see \un below; it's better)
\let\ced=\c         % cedilla
\let\du=\d          % dot-under
\let\um=\H          % Hungarian umlaut
\let\sll=\l         % slashed (suppressed) l (Polish)
\let\Sll=\L         % " L
\let\slo=\o         % slashed o (Scandinavian)
\let\Slo=\O         % " O
\let\tie=\t
% tie-after (semicircle connecting two letters)
\let\br=\u          % breve
        % Also: \`    grave
        %   \'    acute
        %   \v    hacek (check)
        %   \^    circumflex (hat)
        %   \~    tilde (squiggle)
        %   \=    macron (bar-over)
        %   \.    dot (over)
        %   \"    umlaut (dieresis)
        %   \aa \AA   A-with-circle (Scandinavian)
        %   \ae \AE   ligature (Latin & Scandinavian)
        %   \oe \OE   " (French)
        %   \ss   es-zet (German sharp s)
        %   \$  \#  \&  \%  \pounds  {\it\&}  \dots

% Abbreviations for Greek letters

\def\a{\alpha}
\def\b{\beta}
\def\c{\chi}
\def\d{\delta}
\def\e{\epsilon}        % Also, \varepsilon
\def\f{\phi}            %   \varphi
\def\g{\gamma}
\def\h{\eta}
\def\i{\iota}
\def\j{\psi}
\def\k{\kappa}          % Also, \varkappa (see below)
\def\l{\lambda}
\def\m{\mu}
\def\n{\nu}
\def\o{\omega}
\def\p{\pi}         % Also, \varpi
\def\q{\theta}          %   \vartheta
\def\r{\rho}            %   \varrho
\def\s{\sigma}          %   \varsigma
\def\t{\tau}
\def\u{\upsilon}
\def\x{\xi}
\def\z{\zeta}
\def\D{\Delta}
\def\F{\Phi}
\def\J{\Psi}
\def\L{\Lambda}
\def\O{\Omega}
\def\P{\Pi}
\def\Q{\Theta}

\def\sfA{{\sf A}}                       \def\sfa{{\sf a}}
\def\sfB{{\sf B}}           \def\sfb{{\sf b}}
\def\sfC{{\sf C}}           \def\sfc{{\sf c}}
\def\sfD{{\sf D}}           \def\sfd{{\sf e}}
\def\sfE{{\sf E}}           \def\sfe{{\sf f}}
\def\sfF{{\sf F}}           \def\sff{{\sf f}}
\def\sfG{{\sf G}}           \def\sfg{{\sf g}}
\def\sfH{{\sf H}}           \def\sfh{{\sf h}}
\def\sfI{{\sf I}}           \def\sfi{{\sf i}}
\def\sfJ{{\sf J}}           \def\sfj{{\sf j}}
\def\sfK{{\sf K}}           \def\sfk{{\sf k}}
\def\sfL{{\sf L}}           \def\sfl{{\sf l}}
\def\sfM{{\sf M}}           \def\sfm{{\sf m}}
\def\sfN{{\sf N}}           \def\sfn{{\sf n}}
\def\sfO{{\sf O}}           \def\sfo{{\sf o}}
\def\sfP{{\sf P}}           \def\sfp{{\sf p}}
\def\sfQ{{\sf Q}}           \def\sfq{{\sf q}}
\def\sfR{{\sf R}}           \def\sfr{{\sf r}}
\def\sfS{{\sf S}}           \def\sfs{{\sf s}}
\def\sfT{{\sf T}}           \def\sft{{\sf t}}
\def\sfU{{\sf U}}           \def\sfu{{\sf u}}
\def\sfV{{\sf V}}           \def\sfv{{\sf v}}
\def\sfW{{\sf W}}           \def\sfw{{\sf w}}
\def\sfX{{\sf X}}           \def\sfx{{\sf x}}
\def\sfY{{\sf Y}}           \def\sfy{{\sf y}}
\def\sfZ{{\sf Z}}           \def\sfz{{\sf z}}

% Calligraphic letters

\def\ca{{\cal A}}
\def\cb{{\cal B}}
\def\cc{{\cal C}}
\def\cd{{\cal D}}
\def\ce{{\cal E}}
\def\cf{{\cal F}}
\def\cg{{\cal G}}
\def\ch{{\cal H}}
\def\ci{{\cal I}}
\def\cj{{\cal J}}
\def\ck{{\cal K}}
\def\cl{{\cal L}}
\def\cm{{\cal M}}
\def\cn{{\cal N}}
\def\co{{\cal O}}
\def\cp{{\cal P}}
\def\cq{{\cal Q}}
\def\car{{\cal R}}
\def\cs{{\cal S}}
\def\cat{{\cal T}}
\def\cu{{\cal U}}
\def\cv{{\cal V}}
\def\cw{{\cal W}}
\def\cx{{\cal X}}
\def\cy{{\cal Y}}
\def\cz{{\cal Z}}

\def\bd{\begin{displaymath}}
\def\ed{\end{displaymath}}
\def\be{\begin{equation}}
\def\ee{\end{equation}}
\def\bq{\begin{eqnarray}}
\def\eq{\end{eqnarray}}
\def\ba{\begin{array}}
\def\ea{\end{array}}
\def\bqn{\begin{eqnarray*}}
\def\eqn{\end{eqnarray*}}
\def\half{\frac{1}{2}}
\def\st{\star}
\def\v{\varphi}
\def\lbl{\label}
\def\dd{\partial}
\def\rar{\rightarrow}
\def\itmb{\item[$\bullet$]}
\def\half{\frac12}
\def\intin{\int_{-\infty}^{+\infty}}

\newcommand{\ad}[1]{#1^{\dagger}}
\newcommand{\lp}{\left(}
\newcommand{\rp}{\right)}
\newcommand{\nit}{\noindent}
\newcommand{\ct}[1]{\cite{#1}}
\newcommand{\bi}[1]{\bibitem{#1}}
\newcommand{\dg}{\dagger}
\newcommand{\der}[1]{\frac{\partial}{\partial #1}}
\newcommand{\dr}[2]{\frac{\partial #2}{\partial #1}}

\begin{titlepage}
\title{Quantum Potential Approach to Class of Quantum Cosmological
Models}
\author{Arkadiusz B\sll{}aut\thanks{e-mail address:
ablaut@ift.uni.wroc.pl}
and Jerzy Kowalski Glikman\thanks{e-mail address:
jurekk@ift.uni.wroc.pl}\\
Institute for Theoretical Physics\\
University of Wroc\sll{}aw\\
Pl.\ Maxa Borna 9, PL--50-205 Wroc\sll{}aw, Poland}
\maketitle
\begin{abstract}
In this paper we discuss the quantum potential approach of Bohm in the
context of quantum cosmological model. This approach makes it possible
to convert the wavefunction of the universe to a set of equations
describing the time evolution of the universe. Following Ashtekar et.\
al., we make use of quantum canonical transformation to cast a class
of
quantum cosmological models to a simple form in which they can be
solved
explicitly, and then we use the solutions do recover the time
evolution.
\end{abstract}
\end{titlepage}

\section{Introduction}
The problem of the origin of the universe is perhaps as old as the
mankind, but only in the recent years our understanding of the nature
had
become sufficiently deep to address it in the framework of physics.
The
theories we have in hands nowadays make it possible to trace the
evolution
of the
universe starting from tiny fraction of a second after the mysterious
Big
Bang. This is, without doubts, one of the greatest achievements of the
modern science.
However, during these investigations it became more and more clear
that
our present knowledge is not sufficiently developed to answer
questions
concerning the Big Bang itself and what happened before. Actually this
last question may happen to be not well posed, as there are
indications that ``before Big Bang'' time did not exist at all.

This circle of problems is the subject of a newly developed branch of
physics which incorporates recent developments in high energy physics,
quantum field theory, and quantum theory of gravity, and is
called quantum cosmology. Even if in the recent years there is growing
activity in this field, there is a number of problems which not only
are
unsolved, but concerning which there is even no consensus, as to which
of many possible questions are relevant. Let us list some of these
problems.

\begin{itemize}
\itmb It is natural to attack the problems of quantum cosmology by
making use of some hybrid theory constructed from quantum field theory
and the theory of gravity. Such a theory is certainly not known, but
whatever final shape it would take, it will provide us with the class
of
``wavefunctions of the universe'' as solutions. The problem is how
should we interpret such  wavefunctions. In quantum mechanics, a
wavefunction provides us with the probability of finding a system in
some
configuration. But this is not really what we expect of quantum
cosmology.
The problem of major interest is how universe evolved and what is the
class
of initial conditions leading to the universe we observe. One can also
pose a question as to whether there are any alternatives, i.e., are
there any alternative consistent universes.
\itmb It is well known that the fundamental symmetry of general
relativity is the general coordinate invariance. This leads, through
the
Dirac procedure, to the quantum theory which has reparametrization
invariance built-in by default. This means, in particular, that the
theory is not sensitive to any time reperametrizations, and therefore
happens to be completely time-independent. Then how we can talk about
`evolution'? This is the problem of time, which is
probably the most important conceptual problem of quantum gravity
\ct{Isham}.
\itmb Equally well known is the fact that since the equations
governing
quantum mechanical evolution are linear, any linear combination of
their
solutions is a solution itself. But how could we interpret such a
solution consisting of, say, a sum of two different wave functions.
Should we interpret this solution as describing two universes, or
maybe
hope that for some mysterious reasons, only one of them is relevant
for
our observations of the universe? This is the problem of decoherence
(for review, see \ct{Isham}, section 5.5 and \ct{Zurek} and references
therein.)
\end{itemize}

In the present paper we describe the quantum potential approach to
quantum theories proposed by David Bohm \ct{Bohm} and developed in the
context of quantum gravity in \ct{AJ}, and for
cosmology in \ct{JV} and \ct{Vink}. This approach makes
it possible not only to identify trajectories associated with `wave
function of the universe', but also to circumvent both the time and
decoherence problems mentioned above. By making use of the important
technical result of Ashtekar, Tate, and Uggla \ct{ATU1}, \ct{ATU2},
who were able to transform the
Hamiltonian constraint of some interesting cosmological models to the
form of free two or three dimensional wave equation, we are able to
find
a large class of exact wave functions for these models and to identify
the corresponding time evolutions.

Our investigations are undertaken within the framework of
minisuperspace approximation to quantum gravity. This means that out
of
the infinite number of degrees of freedom of the full theory, we
choose to
investigate the dynamics of only finite number of modes. This approach
is frequently criticized because there is no apparent reason why such
a
handicapped theory is to describe the evolution of real universe.
However, one may argue that such a reduced model can result from
taking into account the solutions of the full quantum theory with
 high degree of symmetry, exactly in the same way as one derives the
 Friedmann--Robertson--Walker cosmological model from classical
general
 relativity.
 In fact, there is no apparent reason whatsoever
for the FRW class of solutions of Einstein
equations to describe the physical universe, but surprisingly enough
it
does with great accuracy. There is, no doubt, a great mystery in the
fact that our
(classical) universe is homogeneous and isotropic, but there is no
clear
reason why this high degree of symmetry should not survive
quantization.
If the  quantum universe also possesses a high degree of symmetry,
the minisuperspace approach may be adequate after all.

The plan of the paper is as follows. In section 2 we discuss the
general
features of the quantum potential approach. In section 3 we introduce
the class
of metrics which will be of our interest and derive the form of
classical and quantum Hamiltonian constraints. In the next section,
following \ct{ATU2}, we analyze the canonical transformations which
drastically simplify the form of Hamiltonian constraint, in fact
in all considered cases this constraint reduces to the  two- or
three-dimensional wave equation.

Having solutions of equations is, as it
is well known, not sufficient. One must decide which of the solutions
are `physical'. The simplest criterion is to demand that the resulting
wave function is normalizable. Therefore, in section 5 we  address
the question of what the inner product is. Section 6 is
devoted to description of a simple class of solutions of resulting
equations and to the derivation of the time evolutions corresponding
to the so obtained `wave
functions of the universe.'

\section{The quantum potential}
In this section we will explain how the quantum potential approach
works
in the context of quantum cosmology. The general features of this
approach (in particular, applied to the standard quantum mechanics and
quantum field theory) are discussed in details in \ct{Bohm},
\ct{Berndl}
 and we
recommend the reader to check these papers for general background and
discussion of conceptual issues.

For the cases which will be the object of our investigations in the
sections to come, it is sufficient to consider a model for which the
whole
of quantum dynamics resides in the single equation\footnote{In what
follows all quantum mechanical operators will be written in sans serif
type face: $\sfa$, $\sfb$, \ldots, $\sfA$, $\sfB$ etc.}
\be
\sfH\,\j(x^i) = \lp \half g^{ij}\nabla_i\nabla_j - V(x^i)\rp\,\j(x^i)
=
\lp\half\Box - V(x^i)\rp\, \j(x^i) =
0,\label{ham}
\ee
where $g^{ij}$ may be $x$-dependent. From now on we will call $\j$ the
wave function of the universe. Let us assume that $\j$ has the
following
polar decomposition
\be
\j = R(x^i) \mbox{exp}\lp\frac{i}{\hbar} S(x^i)\rp\label{wf}
\ee
with both $R$ and $S$ real. Inserting (\ref{wf}) into (\ref{ham}), we
obtain two equations corresponding to real and imaginary part,
respectively. These equations read
%\margingpar{Check signs and coefficients}
\be
\ch[S(x)] = \frac{1}{2} g_{ij}\dr{x^i}{S}\dr{x^j}{S} + V(x^i) =
\frac{\hbar^2}{2} \frac{1}{R}\Box R,\label{Re}
\ee
\be
R\Box S + 2 g^{ij}\dr{x^i}{S}\dr{x^j}{R} =0.\label{Im}
\ee
Equation (\ref{Im}) will not concern us anymore. In this paper we
assume
that the wave function $\j$ is  a solution of equation (\ref{ham}),
and thus this equation is
identically satisfied (even though we may not know what the explicit
form
of the wavefunction is.) On the other hand, equation (\ref{Re}) is of
crucial importance. This equation can be used to derive the time
dependence and then serves as the evolutionary equation in the
formalism.

For, let us introduce time $t$ through the following equation
\be
\frac{d x^i}{d t} = g^{ij}\frac{\d \ch[S(x)]}{\d (\partial S
/\partial x^j)}.\label{time}
\ee
This equation defines the trajectory $x^i(t)$ in terms of the phase
of the
wavefunction $S$. Now we can substitute back equation (\ref{time}) to
(\ref{Re}). Assuming that the matrix $g^{ij}$ has the inverse
$g_{ij}$,
we find ($\dot x^i = \frac{d x^i}{d t}$)
\be
\half g_{ij}\dot{x^i}\dot{x^j} + V(x^i) = \frac{\hbar^2}{2}
\frac{1}{R}\Box
R.\label{evol}
\ee
We see therefore that the quantum evolution differs from the classical
one only by the presence of the quantum potential term
$$
-V_{quant}(x^i) = \frac{\hbar^2}{2} \frac{1}{R}\Box
R
$$
on the right hand side of equation of motion. Since we assume that the
wave function is known, the quantum potential term is known as well.

Equation (\ref{evol}) is not in the form which is convenient for our
further investigations. To obtain the desired form we define classical
momenta
$$
p_i =         \frac{\d \ch[S(x)]}{\d (\partial S
/\partial x^i)} = g_{ij}\dot x^j\label{mom}
$$
and cast equation (\ref{evol}) to the form
\be
\ch \equiv \half g^{ij}p_ip_j + V(x^i) - \frac{\hbar^2}{2}
\frac{1}{R}\Box
R = 0.\label{constr}
\ee

We regard $\ch$ as the generator of dynamics acting through the
Hamilton
equations
\bq
\dot p_i &=& - \dr{x^i}{\ch} \nonumber\\
\dot x^i &=& \dr{p_i}{\ch}.                    \label{Hameq}
\eq

Thus we developed the scheme in which we can identify the time
evolution
corresponding to the wavefunction of the universe. This time evolution
is governed by equations (\ref{Hameq}) subject to the constraint for
initial conditions (\ref{constr}). This completes the technical part
of quantum
potential program. However a number of remarks is in order.

\begin{enumerate}
\item   Equation (\ref{Hameq}) with the hamiltonian $\ch$ defined
by (\ref{constr}) is identical with the classical equations of motion
with the only difference that in the hamiltonian the classical
potential
is appended by the quantum term $V_{quant}$ which is a quantity of
order
$\hbar^2$. Observe that in the course of deriving the quantum
potential equations of motion, we did not make any approximations and
therefore the resulting theory is completely equivalent to the
original
quantum system. What makes this approach different from the standard
quantum
mechanics, is that the wave function whose
interpretation is rather obscured, especially in the case of quantum
gravity and quantum cosmology, is now replaced by  trajectories with
well understood physical meaning.
\item The quantum potential interpretation may be used to obtain a
well
defined semi-classical approximation to quantum theory. Indeed, it
can be
said that the system enters the (semi-) classical regime if the
quantum
potential is much smaller than the regular potential term. This
observation has been extensively used in the paper \ct{JV}, where we
were interested in the various mixed regimes (quantum matter and
classical gravity etc.)
\item One of the major adventages of the quantum potential approach is
that it provides one with an affective and simple way of introducing
time even if the system under consideration has a hamiltonian as one
of
the constraints. In particular this approach serves as a possible
route
to final understanding of the problem of time in quantum gravity. We
do
not intend to suggest that this is the only possible approach to this
problem, however, to our taste, this is the best one available now.
\item Related to this is the problem of interpretation of wave
functions which are real. This problem has been a subject of numerous
investigations, but from the point of view of quantum potential the
resolution of it is quite simple. We just say that real wavefunctions
(of
the universe) represent a model without time evolution (and therefore
time) at all. Recall that the standard (time-dependent) Schr\"odinger
equation does not
have any real solutions. It is clear from the formalism: $\dot x$ is
just equal to
zero, so nothing evolves and therefore there is no clock to measure
time. On the other hand, equation (\ref{evol}) means that for the real
wave function the system settles down to the configuration for which
the
total potential (i.e., classical plus quantum) is equal to zero.
\item There is one particular example of equation (1) which
will be important for us later on. Consider the situation when the
classical potential is absent. Since the kinetic term $\Box$ is a
hyperbolic differential operator, the wave function is a sum of
function
corresponding to `retarded' and `advanced' waves, to wit
$$
\j = \j_1(u) + \j_2(v).
$$
Now the quantum potential is defined in terms of the same operator
$\Box$ acting on the modulus of the {\em total} wave function. We see
therefore that if we want the quantum potential not to vanish, we must
keep both components of the wavefunction decomposition above.
\item The definition of time by equation (\ref{time}) is, of course,
not
unique. In fact, we can can use a more general expression
\be
\dot x^i = N(x) \frac{\d \ch[S(x)]}{\d (\partial S
/\partial x^i)} ,
\ee
where, in the case of gravity, the function $N$ is to be identified
with
the lapse function of ADM formalism.
\end{enumerate}

This completes our formal investigations of the quantum potential
approach. Let us now turn to cosmology.

\section{Classical cosmological models}
This section concerns a class of homogeneous cosmological models,
called the
diagonal, intristically multiply transitive models, whose quantum
evolution we will investigate in the following sections. Our
presentation
essentially follows the paper \ct{ATU2} and we recommend the reader to
consult this paper for more details and references.

A spacially homogeneous 4-metric  can be expressed in the form
\be
ds^2 = -N^2(t)\, dt^2 + \sum_{i=1}^3 g_{ii}(\o^i)^2,
\ee
where $N(t)$ is the lapse function and $\o^i$ are some appropriate
1-forms on the 3-space. To make this metric compatible with the
non-evolutionary Einstein equations, $\o^i$ must satisfy certain
conditions whose explicit form will not concern us. Particular forms
of
$\o^i$ are classified and called Bianchi types.

Since the metric coefficients $g_{ii}$ are positive, they can be
represented in the form
$$
g_{ii} = e^{2\b^i}
$$
and it turns out that the following parametrization, due to Misner,
is very convenient
$$
\lp\begin{array}{c}
\b^1\\\b^2\\\b^3\end{array}\rp
=
\lp\begin{array}{ccc}
1&1&\sqrt3\\
1&1&-\sqrt3\\
1&-2&0\end{array}\rp
\lp\begin{array}{c}
\b^0\\\b^+\\\b^-\end{array}\rp,
$$
$$
(x^0, x^+, x^-) = \frac{1}{\sqrt3}(2\b^0-\b^+,-\b^0+2\b^+, \sqrt3\b^-
).
$$
Defining the momenta associated with the variables $x^i$
$$
\{x^i,p_j\}=\d^i_j
$$
and turning to the so called Taub time gauge which fixes the lapse
function $N$ to be
$$
N_T = 12\exp(3\b^0)
$$
we finally arrive at the following simple form of the single remaining
Einstein equation (in phase space)
\bq
\ch &=& \ch_0+ \ch_+ +\ch_- =0,\label{eeq1}\\
\ch_0 &=& -\half p_0^2 + k_0e^{2\sqrt3 x^0},\label{eeq0}\\
\ch_+ &=& \half p_+^2 + k_+e^{-4\sqrt3 x^+},\label{eeq+}\\
\ch_- &=& \half k_- p_0^2 ,\label{eeq-}
\eq
where the coefficients $k_0, k_{\pm}$ for various models Bianchi are
given in the table above.

Before going any further let us make an important remark. As it is
well
known the Einstein lagrangian for cosmological models has the generic
form
$$
\frac{1}{M_P^2}\int dt (p \dot x - N(t)\ch).
$$
Setting  $M_P=1$, as it is usually done,
means that all quantities are scaled by the Planck mass, length etc.
Thus if in the numerical computations some quantity is equal 1, this
quantity is 1 times an appropriate power of $M_P$ and this means that
1
is just the border between the ``classical'' and the ``quantum''
world.

\begin{table}\centering
\begin{tabular}{|c||c|c|c|c|c|c|}
\hline
&I&II&VIII&IX&KS&III\\\hline
$k_0$&0&0&24&-24&-24&24\\
$k_+$&0&6&6&6&0&0\\
$k_-$&1&1&0&0&0&0\\
\hline
\end{tabular}
\caption{Coefficients of hamiltonian constraint for various Bianchi
models}
\end{table}

There is one more model, the so called Bianchi type V model for which
the
scheme is applicable as well; for this model, the $x$ and $\b$
variables
are identical
and the coefficients are given by $k_0=72$, $k_+=0$,
$k_-=1$.

All the models listed above have the important property of being
separable which makes it possible to find a canonical transformation
leading to the potential-free theory.

\section{Classical and quantum canonical transformations}

It is well known that canonical transformations are a very convenient
tool of classical mechanics. If applicable, they are used to simplify
complicated mechanical systems to the ones which are easy to deal
with.
It turns out however that in the case of quantum mechanics our
understanding of canonical transformations is surprisingly poor. To
our
knowledge the only strong result is due to Berezin \ct{Ber} and covers
only linear canonical transformations. Therefore, in a non-linear case
we cannot refer to any theory, but rather we must just check if a
classical canonical transformation can be elevated to the quantum
regime. The problem is that we must make sure that all the
peculiarities of quantum theory (like, for example, operator ordering
problem) do not spoil the classical transformation. This is exactly
what we will do in this section in the context of models described in
section 3.

The generic piece of the hamiltonian constraint can be written
as
$$
\e\half (p^2 + a \exp{b x}),
$$
where $\e=\pm1$, and $a$, $b$ are coefficients which can be read off
from the table in the previous section. The obvious goal will be
therefore to replace this term, after canonical trasformation, by a
square of pure momentum. Thus the canonical transformation must read
\be
P = \sqrt{p^2 + a \exp{2 b x}}.\label{cant}
\ee
To define the new variable $X$ we must solve the equation
$
\{ X, P\} =1,
$
Now  we must distinguish two cases depending on the sign of the
coefficicent $a$ above.
\bq
X&=& \frac{1}{b}\lp\log [-p + \sqrt{p^2+ ae^{2bx}}] -
\log[\sqrt{a}e^{bx}]\rp, \;\;\;\; a>0,\\
X&=& \frac{1}{b}\lp\log [p - \sqrt{p^2+ ae^{2bx}}] -
\log[\sqrt{-a}e^{bx}]\rp, \;\;\;\; a<0           .
\eq
The inverse canonical transformation reads
\bq
\sqrt{a} e^{bx} &=&\frac{P}{\cosh b X}, \;\;\; p=-P\tanh{b X},\;\;\;
a>0\label{18}\\
\sqrt{-a} e^{bx} &=&\frac{P}{\sinh b X}, \;\;\; p=P\coth{b X},\;\;\;
a<0.
\eq
Let us observe that the momentum $P$ defined by (\ref{cant}) is
positive. This does not pose any problems in classical theory, but we
will have to struggle a little with this condition in the quantum case
below. In the paper \ct{ATU2} one can find a detailed discussion of
topology of the phase space of the models after canonical
transformation. It should be stressed that since the canonical
transformations lead to a free theory, this topology encodes all the
nontrivial information concerning the dynamics of the model.

Let us now turn to the quantum picture. As we mentioned above there is
no complete theory of quantum canonical transformations, and all
particular (nonlinear) cases must be discussed separately. In what
follows, we will discuss the case of positive $a$, the negative $a$
case
can be discussed similarly.

We look for the linear transformation which maps wave functions
$\j(x)$ into
wave functions $\J(X)$. In fact the canonical transformation in our
case
is nothing but the transformation to the basis in which the operators
$\sfp_i^2 + a_i\exp{b_i\,\sfx_i}$ are diagonal. It follows from the
basic
principles of quantum mechanics that such a (unitary and linear)
transformation
exists. This transformation must have the property that
solutions of constraint equations before and after the canonical
transformation are mapped into each other. Moreover we would like
this
transformation to be unitary; this, however requires the knowledge of
the
inner products. Since the inner product in the free case is easy to
construct
we use the transformation to {\em define} the inner product in the
space of
original wave functions.

Thus we start with writing
$$
\j(x) = \int^{+\infty}_{-\infty} dX\, <x|X>\, \J(X).
$$
To find the kernel $<x|X>$ satisfying the requirement above we
investigate
the action of the quantum operators corresponding to (\ref{18}) on
both sides
of the above equation. To this end, we must choose some particular
operator
ordering, and it turns out that only the ordering
``$\sfP\sfQ$'' leads to the desired result.

We therefore assume that
\be
\sfp\j(x) = -\int^{+\infty}_{-\infty} dX\, <x|X>\, \sfP\tanh
b\sfX\J(X),
\ee
\be
\sqrt{a}\exp(b\,\sfx)\j(x) = \int^{+\infty}_{-\infty} dX\, <x|X>\,
\sfP\frac{1}{\cosh b\sfX}\J(X),
\ee
{}From these two conditions we have (in position representation)
after integrating by parts
$$
-i\der{x}<x|X> = -i\der{X}\lp <x|X>\rp\tanh bX,
$$
and
$$
\sqrt{a}\exp(bx)<x|X> = i\der{X}\lp <x|X>\rp\frac{1}{\cosh bX},
$$
which can be solved to give
\be
<x|X> = \exp\lp-i\frac{\sqrt a}{b}e^{bx}\sinh(bX)\rp
\label{kernel}
\ee
In general two or three dimensional case the kernel has the form
$$
<x_1,x_2,x_3|X_1,X_2,X_3> = \exp\lp-i\sum_{i=1}^3\frac{\sqrt a_i}{b_i}
e^{bx_i}\sinh(bX_i)\rp.
$$
It should be observed that in order that the procedure described above
make sense and that solutions of constraints are mapped to solutions,
the wave function $\J(X)$ must be integrable with the kernel
(\ref{kernel}) at least together with its first derivative.
By changing variables in the integral $\sinh bX = Y$,
we see that
$$
\j(x) = \int^{+\infty}_{-\infty} dY\,
\exp\lp-i\frac{\sqrt a}{b}e^{bx}\,Y\rp\frac{1}{b\sqrt{1+ Y^2}}\J(X(Y))
$$
and the integral exists (in the sense of principal value) if $\J(Y)$
tends to a constant when
$Y\rightarrow\pm\infty$. Of course, $\J(Y)$ cannot have any nasty
singularities in finite interval.

Therefore given any solution of simple free system we can, in
principle,
construct a solution of the original theory. The problem is that the
integral transform defined above is very complex because the kernel
oscillates very rapidly for large $X$. For that reason it is not only
hopeless to find the integral in terms of tabulated functions but it
is
even
difficult to compute the transform numerically. However some
approximate
methods (for large $x$, for example) may work for some $\J(X)$.

\section{The inner product}

We succeeded therefore to reduce our problem to the problem of free
scalar particle in two or three dimensions. However not all solutions
of
the constraint equation
\be
\Box\J(X)=0\label{1}
\ee
can be interpreted as a physical wavefunction. First of all, in order
to
make contact with the original theory, we must make sure that the
integral transform defined in the previous section exists. Secondly,
the
wavefunction is supposed to describe physical probabilities, thus it
must be normalizable. Thus we must define an inner product in the
space
of solutions  of equation (\ref{1}).

The problem of defining the inner product for Dirac quantization
scheme is unsolved in general (see however \cite{inner}). In the case
in
hand, however, it suffices just to construct the inner product.

To this end, let us observe that any solution of equation (\ref{1})
can
be, in momentum representation, written in the form
$$
\J(P) = \d(\h^{ij}P_iP_j)\tilde{\J}(P),
$$
where $\tilde{\J}(P)$ is an arbitrary function of $P$ sufficiently
well behaving at $\h^{ij}P_iP_j=0$.
In any other representation one must replace $\d(\h^{ij}P_iP_j)$ with
the delta function of the corresponding operator
\be
\J = \d(\h^{ij}\sfP_i\sfP_j)\tilde{\J} \equiv \int_{-
\infty}^{+\infty} d\l\,
\exp(i\l \h^{ij}\sfP_i\sfP_j)\tilde{\J}.
\ee
Now we can define the inner product on the space of solutions of
costraint equation to be
$$
\parallel \J\parallel^2 = \int
\tilde{\J}^{\star}\d(\h^{ij}\sfP_i\sfP_j)
\tilde{\J}.
$$
This inner product is not very convenient for our future
investigations.
The reason is that the wave function being a solution of (two
dimensional) wave operator has a general form
$$
\J(X_0,X_1) = \intin dkA_ke^{ikU} + \intin dlB_l e^{ilV},
$$
where $U = X_0+X_1$ and $V=X_0-X_1$. For that reason we would like to
have an
equivalent
definition of the inner product for wave functions in position
representation.

Following \ct{inner} we proceed as follows. We define the inner
product
to be
$$
\parallel \J(X)\parallel^2 = \intin d^2X \J^{\st}(X)\hat{\m}\J(X)
,$$
where $\hat{\m}$ is a gauge fixing operator. In the case in hand this
operator is chosen to be $\frac{\sfX_0}{2\sfP_0}$.\footnote{In
general case
when the comutator of gauge fixing operator and gauge constraint  is
not
constant, the formula above must be generalized, see \ct{inner}.}
This operator acts in a simple way in momentum representation. We have
therefore\footnote{$<\;|\;>$ is the inner product of the original
Hilbert (or, more generally Krein) space of the model. Thus $<P|X>=
e^{iPX}$.}
\bqn
&&\parallel \J(X)\parallel^2 =\\
&=& \intin d^2X'd^2Pd\l \J^{\st}(X')<X'|P>\exp\lp-\l\frac{1}{2P_0}
\der{P_0}\rp <P|X>\J(X) =\\
&=& \intin d^2X'd^2Pd\l \J^{\st}(X')e^{-i(X'_0P_0+X'_1P_1)}
e^{i((X_0(P_0-\frac{\l}{2P_0})+X_1P_1)}\J(X_0,X_1).
\eqn
Integrating over $P_1$ and $X'_1$, we get
$$
\intin dX'_0dX_1dX_0dP_0d\l \J^{\st}(X'_0,X_1)e^{-iX'_0P_0}
e^{i(X_0(P_0-\frac{\l}{2P_0})}\J(X_0,X_1)=
$$
$$
\intin dX'_0dX_1dX_0dP_0 \J^{\st}(X'_0,X_1)e^{iP_0(X_0-X'_0)}
\d\lp\frac{X_0}{2P_0}\rp\J(X_0,X_1),
$$
which after integration over $X_0$ gives
$$
\intin dX'_0dX_1dP_0 \J^{\st}(X'_0,X_1)e^{-iP_0X'_0}
{2P_0}\J(0,X_1)=
$$
$$
-2\intin dX_1\left.\der{X'_0} \J^{\st}(X'_0,X_1)\right|_{X'_0=0}
\J(0,X_1).
$$
This inner product can be also written as
\be
\intin dX\left.
\J^{\st}(Y,X)\stackrel{\leftrightarrow}
{\partial_Y}\J(Y,X)\right|_{Y=0}
\lbl{in1}
\ee
This is in fact the standard inner product for massless spin zero
particle, as it could have been expected. Observe that we clearly
have here
normalizable negative norm states. This fact is well known, and there
is
an open problem what is the physical meaning of such states, and/or if
their existance indicate the need of third quantization.

\section{A simple model}
In this section we make use of the machinery built above to discuss a
simplest possible nontrivial model, namely
\be
\J(X_0,X_1) = e^{i(k+l)U}+e^{i(k-l)V},\label{wf1}
\ee
where $U,V = X_0\pm X_+$ as before.
The above wave function is certainly not physical (it is not
normalizable), however it has the virtue to analogous to the plane
wave
states in the standard quantum field theory. For that reason all the
results below should be taken with the grain of salt, however it does
not seem unlikely that some physical prediction of this simple model
may
hold for the full theory which will be discussed elsewhere.

Given the wavefunction (\ref{wf1}), one can readily find the quantum
potential. Equation (7) reads in our case
$$
-P^2_0+P^2_+ + (k^2-l^2) =0
$$
and from the hamiltonian equation of motion we find easily that
\bq
P_0 =k && X^0 = -kt\\
P_+ = l && X^+ = lt + t_0
\eq
{}From the condition for the canonical transformation to be meaningful,
we
see that both $k$ and $l$ must be positive. Now we are ready to turn
to
the starting variables. To this end we choose to work with the Bianchi
type IX model, and using equation (\ref{18}), we find
\bq
\sqrt{48} e^{2\sqrt{3}x^0} &=&\frac{k}{\cosh2\sqrt{3} kt}, \;\;\; p_0
=-l\tanh{2\sqrt{3} kt},\label{e1}\\
\sqrt{12} e^{-4\sqrt{3}x^+} &=&\frac{l}{\cosh 4\sqrt{3} (lt+t_0)},
\;\;\; p_+=l\tanh{4\sqrt{3} (lt+t_0)}.\label{e2}
\eq
Looking at the Misner parametrization, we see that the metric depends
on
$t$ through the combination \footnote{In wiev of construction of
sect.\ 2,
it is not surprising that the trajectories below are integral curves
of
the Wheeler - De Witt current \ct{Vilenkin}. We would like to thank
the
anonymous referee for pointing this fact to us.}
\bqn
\b^0 &=& \frac{1}{3}\log\lp
\frac{k}{4\sqrt3}\lp\frac{2\sqrt3}{l}\rp^{1/4}
\frac{\cosh^{1/4}(4\sqrt3lt+t_0)}{\cosh(2\sqrt3 kt)}\rp\\
\b^+ &=& \frac{1}{6}\log\lp
\frac{k}{2l}
\frac{\cosh(4\sqrt3lt+t_0)}{\cosh(2\sqrt3 kt)}\rp
\eqn
Now we can discuss the characteristic features of the model. It is
described by the metric
$$
ds^2 = - N_T^2 dt^2 + e^{2\b^0}\lp
e^{2\b^+} (\o^1)^2 +
e^{2\b^+} (\o^2)^2 +
e^{-4\b^+} (\o^3)^2 \rp,
$$
where $e^{2\b^0}$ is the scale factor and $e^{2\b^+}$ is a measure of
the anisotropy. The physical time is given by the monotoneous function
of the parameter $t$,
$$
\t \sim \int^{\t}
\frac{\cosh^{1/4}(4\sqrt3lt+t_0)}{\cosh(2\sqrt3 kt)}\, dt
$$
and for large $t$, $\t$ behaves as $\t \sim \frac{1}{l-2k}e^{\sqrt3
t(l-2k)}$. The scale factor is a symmetric function of $t$ with one
extremum, which is a minimum for $l>2k$, and a maximum for $l<2k$.
Similarly, the anisotropy factor has only one maximum for $2l>k$ and
only one minimum for $2l<k$. Thus we have four qualitatively different
classes of models, depending on different values of $k,l$. Observe
that
the classical behaviour corresponds to $k=l$ (in this case the quantum
potential vanishes.)

The model above should be viewed only as an illustration of
application
of our method to quantum cosmological models. As it was mentioned at
the
beginning of this section its physical value is diminished by the fact
that the corresponding wavefunction is not normalizable. The more
realistic models are currently under consideration and the results
will be published in a separate paper.
\clearpage

\newcommand{\prev}[3]{Phys.\ Rev.\ {\bf {#1}},  {#2}, ({#3})}
\newcommand{\prep}[3]{Phys.\ Rep.\ {\bf {#1}},  {#2}, ({#3})}
\newcommand{\prevD}[3]{Phys.\ Rev.\  {\bf D {#1}},  {#2}, ({#3})}
\newcommand{\pletB}[3]{Phys.\ Lett.\  {\bf  {#1} B},  {#2}, ({#3})}
\newcommand{\nphB}[3]{Nucl.\ Phys.\  {\bf B {#1}},  {#2}, ({#3})}
\newcommand{\jstp}[3]{J.\ Stat.\ Phys.\  {\bf {#1}},  {#2},  ({#3})}
\newcommand{\scam}[3]{Scientific American {\bf {#1}},  {#2},  ({#3})}

\end{document}